\documentclass[12pt]{article}

     \newlength{\dinwidth}
     \newlength{\dinmargin}
\def\Journal#1#2#3#4{{#1} \textbf{#2}, {#3} (#4)}
\def\EPJC{\textit{Eur.\ Phys.\ J.} C}
\def\NIMA{\textit{Nucl.\ Instrum.\ Meth.} A}

\def\PRD{\textit{Phys.\ Rev.} D}
\def\ZPC{\textit{Z. Phys.} C}

\usepackage{cite}
\usepackage{amssymb}
\usepackage{graphicx}

\newcommand{\Angle}{\measuredangle}

\newcommand{\dd}{\mathrm{d}}
\newcommand{\GeV}{\ensuremath{\;\mathrm{GeV}}}
\newcommand{\Eep}{E_e{\!\!'}\,}

\newcommand{\Li}{\mbox{Li}_2}

\begin{document}
\vspace*{10mm}
\begin{center}  \begin{Large} \begin{bf}
    QED corrections to DIS with a tagged photon at HERA\\
  \end{bf}  \end{Large}
  \vspace*{5mm}
  \begin{large}
     H. Anlauf$^a$%
     \footnote{Supported by Bundesministerium f\"ur Bildung, Wissenschaft,
               Forschung und Technologie (BMBF), Germany.}%
     ,
     A.B. Arbuzov$^{bc}$, E.A. Kuraev$^c$\\
  \end{large}
\end{center}
$^a$ Fachbereich Physik, Siegen University, 57068 Siegen, Germany\\ 
$^b$ University of Turin and INFN group, 10125, Turin, Italy\\
$^c$ Joint Institute for Nuclear Research, 141980, Dubna, Russia\\
\begin{quotation}
\noindent
\textbf{Abstract:}
    We report on the calculation of the QED radiative corrections to deep
    inelastic scattering with a tagged photon with next-to-leading logarithmic
    accuracy.  Numerical results are given for different experimental setups
    for the case of the HERA collider.
\end{quotation}


\section{Introduction}
\label{sec:Intro}

One of the major tasks of the experiments at the HERA $ep$ collider is the
determination of the structure functions of the proton, $F_2(x,Q^2)$ and
$F_L(x,Q^2)$, over a broad range of the kinematical variables.  Especially the
extension of the measurements from fixed target experiments to the range of
small Bjorken $x < 10^{-4}$ and for $Q^2$ of a few GeV$^2$ is of particular
interest, as it provides a challenge for theoretical attempts to understand
the details of the dynamics of quarks and gluons inside the nucleon.

There exist several ways to separately extract $F_L(x,Q^2)$ and $F_2(x,Q^2)$
from the experimental data.  Besides indirect methods that rely on
extrapolations or QCD fits (see e.g., \cite{DIS99:Arkadov}), there exist two
direct ones.  The first requires to run the collider at different
center-of-mass energies.  The second method was suggested by Krasny et
al.~\cite{KPS92} and utilizes radiative events with an exclusive hard photon
registered in the forward photon detector (PD).  Such a device is actually
part of the luminosity monitoring system of the H1 and ZEUS experiments.  The
idea of this method is that emission of photons in a direction close to the
incoming electron corresponds to a reduced effective beam energy.  This
effective beam energy for each radiative event is determined from the energy
of the hard photon observed (tagged) in the PD.

The potential of this method is supported by recent preliminary results from
the H1 collaboration of an analysis for $F_2$ \cite{H1:ISR-prelim} (for
earlier analyses that did not take into account QED radiative corrections see
\cite{H1:rad,ZEUS96}).  The feasibility of a determination of $F_L$ was
studied in \cite{FGMZ96}.

It is the purpose of this contribution to review recent calculations of the
QED radiative corrections \cite{AAKM:ll,AAKM:JETP,AAKM:nlo,Anl99:Sigma} to the
process under consideration.


\section{Kinematics and Lowest Order Cross Section}

A convenient set of kinematic invariants that takes into account the energy
loss from the collinearly radiated photon in the process
\begin{equation} \label{eq:tag-kin-born}
  e(p) + p(P) \to e(p') + \gamma(k) + X(P'),
\end{equation}
is given by \cite{KPS92}:
\begin{equation}
\label{eq:kin-vars}
        \hat{Q}^2 =
        -(p-p'-k)^2 \, , \quad
        \hat{x}   =
        \frac{\hat{Q}^2}{2P \cdot (p-p'-k)} \, , \quad
        \hat{y}   =
        \frac{P \cdot (p-p'-k)}{P \cdot (p-k)} \, .
\end{equation}
We restrict ourselves to the case where the polar angle $\vartheta_\gamma$ of
the photon (measured with respect to the incident electron beam) is assumed to
be very small, $\vartheta_\gamma \leq \vartheta_0$, with $\vartheta_0$ being
about 0.45~mrad in the case of the PD of H1.  Therefore, it is emitted almost
collinearly.  In addition we require that the angle $\theta$ of the final
electron be large compared to the angle $\vartheta_0$.

It is convenient to denote by $z$ the energy fraction of the electron after
initial state radiation of a tagged collinear photon,
\begin{equation} \label{def:z}
  z = \frac{2P \cdot (p - k)}{S} = \frac{E_e - E_\gamma}{E_e}
    = \frac{\hat{Q}^2}{\hat{x}\hat{y} S} \, ,
  \qquad \mbox{with} \quad S = 2 P \cdot p \, ,
\end{equation}
where $E_e$ is the electron beam energy, and $E_\gamma$ represents the energy
deposited in the forward PD.

The Born cross section, integrated over the solid angle of the photon detector
($0 \leq \vartheta_\gamma \leq \vartheta_0, \; \vartheta_0 \ll \theta$) takes
a factorized form (see also \cite{KPS92,BKR97,AAKM:ll,AAKM:JETP,AAKM:nlo}):
\begin{equation} \label{eq:Born}
  \frac{1}{\hat{y}} \,
  \frac{\dd^3\sigma_\mathrm{Born}}{\dd\hat{x}\,\dd\hat{y}\,\dd z} =
  \frac{\alpha}{2\pi} \, P(z,L_0) \, \tilde\Sigma(\hat{x},\hat{y},\hat{Q}^2) \, ,
\end{equation}
where
\begin{eqnarray} \label{eq:Sigma-tilde}
  \tilde\Sigma(\hat{x},\hat{y},\hat{Q}^2)
  & = & \frac{2\pi\alpha^2(-\hat{Q}^2)}{\hat{Q}^2\hat{x}\hat{y}^2}
  \left[ 2(1-\hat{y}) - 2\hat{x}^2\hat{y}^2\frac{M^2}{\hat{Q}^2}
    + \left( 1+4\hat{x}^2\frac{M^2}{\hat{Q}^2}\right) \frac{\hat{y}^2}{1+R}
  \right]
  F_2(\hat{x},\hat{Q}^2) ,
  \nonumber
\end{eqnarray}
with
\begin{eqnarray}
  P(z,L_0) & = & \frac{1+z^2}{1-z} L_0 - \frac{2z}{1-z} \, , \qquad
  L_0 = \ln\left(\frac{E_e^2\vartheta_0^2}{m^2}\right)     ,
  \qquad
  \alpha(-\hat{Q}^2) = \frac{\alpha}{1-\Pi(-\hat{Q}^2)} \, , \nonumber \\
  \label{eq:R}
  R & = & R(\hat{x},\hat{Q}^2) =
  \left( 1+4\hat{x}^2\frac{M^2}{\hat{Q}^2} \right)
  \frac{F_2(\hat{x},\hat{Q}^2)}{2\hat{x}F_1(\hat{x},\hat{Q}^2)} \, - 1
  \, .
\end{eqnarray}
The quantities $F_2$ and $F_1$ denote the proton structure functions, $M$ and
$m$ are the proton and electron masses.  We explicitly include the correction
from vacuum polarization $\Pi(-\hat{Q}^2)$ in the virtual photon propagator,
and we neglect the contributions from Z-boson exchange and $\gamma$-Z
interference, because we are interested mostly in the kinematic region of
small momentum transfer $\hat{Q}^2$.


\section{Radiative Corrections}
\label{sec:rc}

The cross section (\ref{eq:Born}) describes the process
(\ref{eq:tag-kin-born}) to lowest order in perturbation theory.  We shall
restrict our discussion to the model-independent corrections to the electron
line.  The radiative corrections to this cross section are composed of
contributions by corrections due to virtual photon exchange, soft photon
emission, and emission of a second hard photon, with one (or both) of the hard
photons being tagged in the PD.  Because of its coarse granularity, we shall
assume that the PD cannot measure photons individually but only their total
energy when two hard photons simultaneously hit the PD in different locations.

As is well known (see e.g., the discussion in \cite{Wol97}), radiation of an
(additional) hard photon influences the experimental determination of the
kinematical variables.  Therefore, the calculation of the contributions from
the emission of two hard photons will depend on the chosen method.  On the
other hand, the virtual corrections and the contributions from soft photon
emission in addition to the hard one will be independent of this choice.


\subsection{Virtual and Soft Corrections}

The virtual and soft corrections to the lowest order cross section were
obtained in \cite{AAKM:JETP,AAKM:nlo}.  Since the tagged photon is emitted
almost collinearly, the cross section takes again a simple, factorized form,
\begin{equation} \label{eq:V+S}
  \frac{1}{\hat{y}} \,
  \frac{\dd^3\sigma_\mathrm{V+S}}{\dd\hat{x}\,\dd\hat{y}\,\dd z} =
  \frac{\alpha^2}{4\pi^2} \left[ P(z,L_0) \tilde{\rho} - T \right]
  \tilde\Sigma(\hat{x},\hat{y},\hat{Q}^2) \, ,
\end{equation}
with
\begin{eqnarray} \label{eq:rho-tilde}
  \tilde{\rho} & = &
  2(L_Q - 1)\ln\frac{\Delta^2}{Y}
  + 3L_Q + 3\ln z - \ln^2Y - \frac{\pi^2}{3} - \frac{9}{2}
  + 2\Li \left( \frac{1+c}{2} \right) ,
  \nonumber \\
  T & = &
  \frac{1+z^2}{1-z}(A\ln z + B)
  - \frac{4z}{1-z}L_Q\ln z
  - \frac{2-(1-z)^2}{2(1-z)} L_0 + {\mathcal O}({\mathrm{const}}) \, ,
  \nonumber \\
  A & = & {} -L_0^2 + 2L_0L_Q - 2L_0\ln(1-z) \, ,
  \quad
  B   =   \left[ \ln^2 z - 2 \Li (1-z)\right] L_0 \, ,
  \nonumber \\
  L_Q & = & \ln\frac{\hat{Q}^2}{z m^2} \, , \qquad
  \Li (t) = -\int\limits_{0}^{t}\frac{\dd u}{u}\ln(1-u) \, .
\end{eqnarray}
Here $\Delta$ denotes the infrared cutoff for the emission of a soft photon in
addition to the hard one, i.e., $E_{\mathrm{soft}} \leq \Delta \, E_e$,
\begin{equation} \label{eq:Y-c-elastic}
  Y \equiv \frac{\Eep}{E_e}
  = z(1 - \hat{y}) + \hat{x}\hat{y} \, \frac{E_p}{E_e} \; ,
  \quad \mbox{and} \quad
  c \equiv \cos\theta =
  \frac{z(1 - \hat{y}) E_e - \hat{x}\hat{y} E_p}
       {z(1 - \hat{y}) E_e + \hat{x}\hat{y} E_p}
  \; .
\end{equation}

It should be noted that in (\ref{eq:V+S},\ref{eq:rho-tilde}) and also in the
contributions from double hard brems\-strahlung we retain only terms with
double or single large logarithms of the small electron mass $m$, i.e., terms
of order $\alpha^2 L^2$ and $\alpha^2 L$, with $L$ being one of $L_0$ or
$L_Q$.  As the lowest order cross section (\ref{eq:Born}) is of order $\alpha
L$ relative to the DIS cross section, we denote the terms of order $\alpha^2
L^2$ as leading (LL) and those of order $\alpha^2 L$ as next-to-leading
logarithmic (NLL) ones.


\subsection{Double Hard Bremsstrahlung}

Besides the soft and virtual corrections to the lowest order process, we have
to consider also the corrections from hard bremsstrahlung, which in the
present case corresponds to double hard bremsstrahlung.

In the calculation of the contributions from the emission of two hard photons,
it is convenient to decompose the phase space into three regions
\cite{AAKM:nlo}: \textit{i)} both hard photons hit the forward photon
detector, i.e., both are emitted within a narrow cone around the electron beam
$(\vartheta_{1,2} \leq \vartheta_0, \, \vartheta_0 \ll 1)$; \textit{ii)} one
photon is tagged in the PD, while the other is collinear to the outgoing
electron $(\vartheta_2' \equiv \Angle(\vec{k}_2,\vec{p}{\,'}) \leq
\vartheta'_0)$; and finally \textit{iii)} the second photon is emitted at
large angles (i.e., outside the defined narrow cones) with respect to both
incoming and outgoing electron momenta.  The last kinematic domain is denoted as
the semi-collinear one.  For the sake of simplicity, we shall always assume
that $m/E_e \ll \vartheta'_0 \ll 1$.

The contribution from the kinematic region \textit{i)}, with both hard photons
being tagged, and only the sum of their energies measured, does not depend on
the determination of the kinematic variables and reads
\begin{eqnarray} \label{eq:sig-i}
  \frac{1}{\hat{y}} \,
  \frac{\dd^3 \sigma^{\gamma\gamma}_i}{\dd\hat{x}\,\dd\hat{y}\, \dd z}
  & = &
  \frac{\alpha^2}{8\pi^2} L_0
  \Biggl[ L_0 \left( P^{(2)}_{\Theta}(z) +
          2\frac{1+z^2}{1-z}\left(\ln z-\frac{3}{2}-2\ln\Delta\right)\right)
  \nonumber \\
  && \qquad \quad
  {} + 6(1-z) + \left(\frac{4}{1-z}-1-z\right)\ln^2z - 4\frac{(1+z)^2}{1-z}
  \ln\frac{1-z}{\Delta}\Biggr] \tilde\Sigma(\hat{x}, \hat{y}, \hat{Q}^2)
  \nonumber\\ &&
  {} + {\mathcal O}({\mathrm{const}}) \, ,
\end{eqnarray}
where $P^{(2)}_{\Theta}(z)$ can be found in, e.g., \cite{AAKM:JETP,AAKM:nlo}.


In contrast to the above contributions, the contributions from the regions
\textit{ii)} and \textit{iii)} depend on the experimental determination of the
kinematic variables and on the experimental selection of the events, as
discussed below.


Within the required logarithmic accuracy it can be shown that the contribution
from the semi-collinear region \textit{iii)} factorizes as
\begin{equation} \label{eq:sig-iii}
  \frac{1}{\hat{y}} \,
  \frac{\dd^3 \sigma^{\gamma\gamma}_{iii}}{\dd\hat{x}\,\dd\hat{y}\, \dd z}
  =
  \frac{\alpha^2}{\pi^2} P(z,L_0)
  \int\frac{\dd^3 k_2}{|\vec{k}_2|}
  \, \frac{\alpha^2(Q_h^2)}{Q_h^4}
  \, I^{\gamma}(zp,p',k_2)
  \, ,
\end{equation}
with the precise form of the radiation kernel $I^{\gamma}$ depending on the
determination of the kinematic variables; for details we refer to
\cite{AAKM:nlo,Anl99:Sigma}.
In the phase space integration over the hard photon, it is understood that the
angular part of the $k_2$-integration is clearly restricted to the kinematic
region \textit{iii)}, i.e., the full solid angle with the exception of the
separately treated cones around the incoming and outgoing electron.

Finally we turn to the kinematic region \textit{ii)}.  As discussed in
\cite{AAKM:nlo}, the contribution of this region to the observed cross section
depends on the experimental event selection, i.e., on the method of
measurement of the scattered particles.  We shall focus on two scenarios.  The
first one is denoted as an \textit{exclusive} (or bare) event selection, as
only the scattered electron is measured; the hard photon that is emitted
almost collinearly (i.e., within a small cone with opening angle
$2\vartheta_0'$ around the momentum of the outgoing electron) remains
undetected or is not taken into account in the determination of the kinematic
variables.  The second case is a \textit{calorimetric} event selection, when
only the sum of the energies of the outgoing electron and photon is actually
measured if the photon momentum lies inside a small cone with opening angle
$2\vartheta_0^{'}$ along the direction of the final electron.

For the exclusive event selection, when only the scattered electron is
detected, we obtain
\begin{eqnarray} \label{eq:sig-ii-excl}
  \frac{1}{\hat{y}} \,
  \frac{\dd^3 \sigma^{\gamma\gamma}_{ii,\mathrm{excl}}}{\dd\hat{x}\,\dd\hat{y}\, \dd z}
  & = &
  \frac{\alpha^2}{4\pi^2} P(z,L_0)
  \int\limits_{\zeta_\mathrm{min}}^{\zeta_\mathrm{max}}
  \frac{\dd \zeta}{\zeta^2}
  \left[ \frac{1+\zeta^2}{1-\zeta} \left(\widetilde L-1\right) + (1-\zeta) \right]
  \tilde\Sigma_f(x_f,y_f,Q_f^2) \, ,
\end{eqnarray}
where $\tilde\Sigma_f(x_f,y_f,Q_f^2)$ is an implicit function of $\zeta$ via
the relation between the ``internal'' kinematic variables $x_f,y_f,Q_f^2$ and
the ``external'' ones $\hat{x}, \hat{y}, \hat{Q}^2$ (see
refs.~\cite{AAKM:JETP,AAKM:nlo,Anl99:Sigma} for more details).  Also, the
large logarithm $\widetilde L$ generally depends on $\zeta$.  The integration
limits explicitly depend on the chosen determination of kinematic variables.

In the case of a calorimetric event selection, where only the sum of the
energies of the outgoing electron and collinear photon is measured, the
corresponding contribution reads
\begin{eqnarray} \label{eq:sig-ii-cal}
  \frac{1}{\hat{y}} \,
  \frac{\dd^3 \sigma^{\gamma\gamma}_{ii,\mathrm{cal}}}{\dd\hat{x}\,\dd\hat{y}\, \dd z}
  & = &
  \frac{\alpha^2}{4\pi^2} P(z,L_0)
  \int\limits_0^{\zeta_\mathrm{max}}
  \!\! \dd \zeta
  \left[ \frac{1+\zeta^2}{1-\zeta}
    \left(L_0' - 1 + 2\ln\zeta \right) + (1-\zeta) \right]
  \tilde\Sigma(\hat{x}, \hat{y}, \hat{Q}^2) \; ,
\end{eqnarray}
where
$L_0'=\ln\left(\Eep^2\vartheta_0'{}^2/m^2\right)$
is a large logarithm that now depends on the resolution parameter
$\vartheta_0'$ \cite{AAKM:JETP,AAKM:nlo,Anl99:Sigma}.

The total contribution from QED radiative corrections is finally found by
adding up (\ref{eq:V+S}), (\ref{eq:sig-i}), (\ref{eq:sig-iii}), and, depending
on the chosen event selection, (\ref{eq:sig-ii-excl}) or
(\ref{eq:sig-ii-cal}).  The unphysical IR regularization parameter $\Delta$
cancels in the sum, as it should.

It is important to note that the angle $\vartheta_0'$ plays only the r\^{o}le
of an intermediate regulator for the bare event selection and therefore drops
out in the final result.  In the calorimetric case there are no large
logarithmic contributions from final state radiation as long as $\vartheta_0'$
does not become too small.  For more details see \cite{AAKM:nlo}.


\section{Numerical Results}
\label{sec:results}

In this section we shall present numerical results obtained for the leading
and next-to-leading radiative corrections.  As input we used
\begin{equation}
  E_e = 27.5 \GeV \, , \quad
  E_p = 820  \GeV \, , \quad
  \vartheta_0 = 0.5 \; \mathrm{mrad} \, .
\end{equation}
We chose the ALLM97 parameterization \cite{ALLM97} as structure function with
$R=0$, no cuts were applied to the phase space of the second photon, and we
assumed a calorimetric event selection.  For the sake of simplicity we took a
fixed representative angular resolution of $\vartheta_0' = 50 \;
\mathrm{mrad}$ for the electromagnetic calorimeter to separate nearby hits by
a scattered electron and a hard photon, which is close to realistic for the H1
detector at HERA.  Also we disregard any effects due to the magnetic field
bending the scattered charged electron away from a collinear photon.

Below we shall consider two methods that are used for the kinematic
reconstruction of the radiative events and the determination of the kinematic
variables: the electron method and the $\Sigma$ method \cite{BB95}.


\begin{figure}[ht]
  \begin{center}
    \begin{picture}(120,120)
      \put(0,0){
        \includegraphics[width=120mm]{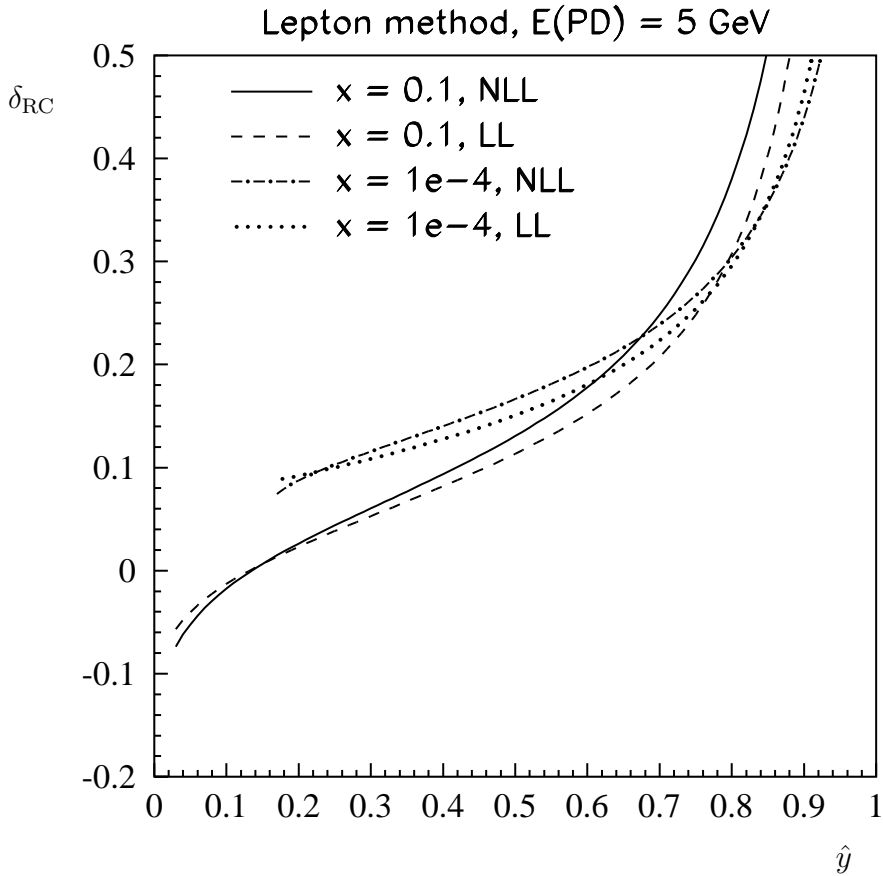}
        }
      \put(104,0){$\hat{y}$}
      \put(-6,101){$\delta_\mathrm{RC}$}
    \end{picture}
    \caption{\textit{Radiative corrections $\delta_\mathrm{RC}$
      (\protect\ref{eq:delta}) with leading and next-to-leading logarithmic
      accuracy at $\hat{x} = 0.1$ and $\hat{x} = 10^{-4}$ and a tagged photon
      energy of 5\GeV\/ for the electron method.  No cuts have been applied to
      the phase space of the second (semi-collinear) photon.}}
    \label{fig:plot1}
  \end{center}
\end{figure}


Figure~\ref{fig:plot1} compares the radiative correction
\begin{equation} \label{eq:delta}
  \delta_\mathrm{RC} =
  \frac{\dd^3\sigma}{\dd^3\sigma_\mathrm{Born}} - 1
\end{equation}
calculated with leading and next-to-leading logarithmic accuracy for the
electron method at $\hat{x}=0.1$ and $\hat{x}=10^{-4}$ and for a tagged energy
of $E_\mathrm{PD} = 5\GeV$.  Similar to the well-known QED corrections for DIS
(see e.g., \cite{BBC+97} and references cited therein), the corrections are
large and positive for large $\hat{y}$, while they are large and negative for
$\hat{y} \to 0$ at large $\hat{x}$.


\begin{figure}[ht]
  \begin{center}
    \begin{picture}(120,120)
      \put(0,0){
        \includegraphics[width=120mm]{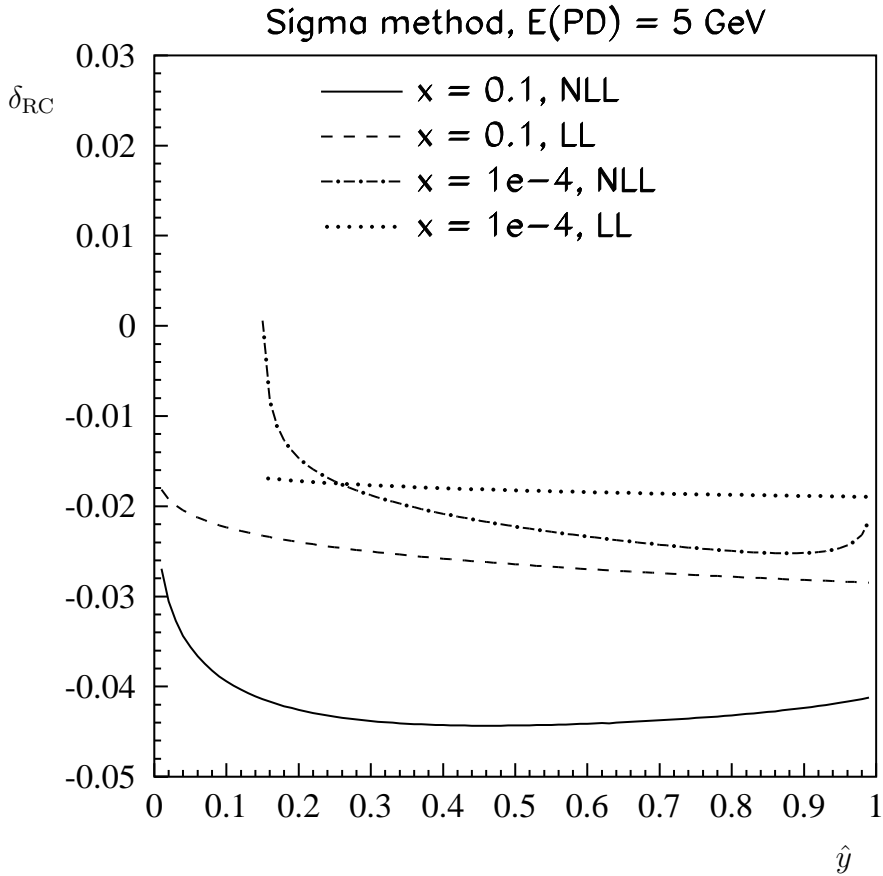}
        }
      \put(104,0){$\hat{y}$}
      \put(-6,101){$\delta_\mathrm{RC}$}
    \end{picture}
    \caption{\textit{Radiative corrections at $\hat{x} = 0.1$ and $\hat{x} =
      10^{-4}$ and a tagged photon energy of 5\GeV\/ for the $\Sigma$ method.
      As is figure \protect\ref{fig:plot1}, no cuts have been applied.}}
    \label{fig:plot2}
  \end{center}
\end{figure}

Figure~\ref{fig:plot2} shows the corresponding corrections for the case of the
determination of the kinematic variables using the $\Sigma$ method
\cite{Anl99:Sigma}.  In contrast to the electron method the corrections appear
to be rather small, being only of the order of 5\%, which can be easily traced
back to the weak dependence of the kinematic variables on undetected initial
state radiation.  This leads to an almost complete cancelation of the leading
logarithmic contributions to the corrections.  On the other hand, the pure
next-to-leading logarithmic parts of the corrections turn out to contribute
significantly due to this suppression.


\begin{figure}[ht]
  \begin{center}
    \begin{picture}(120,120)
      \put(0,0){
        \includegraphics[width=120mm]{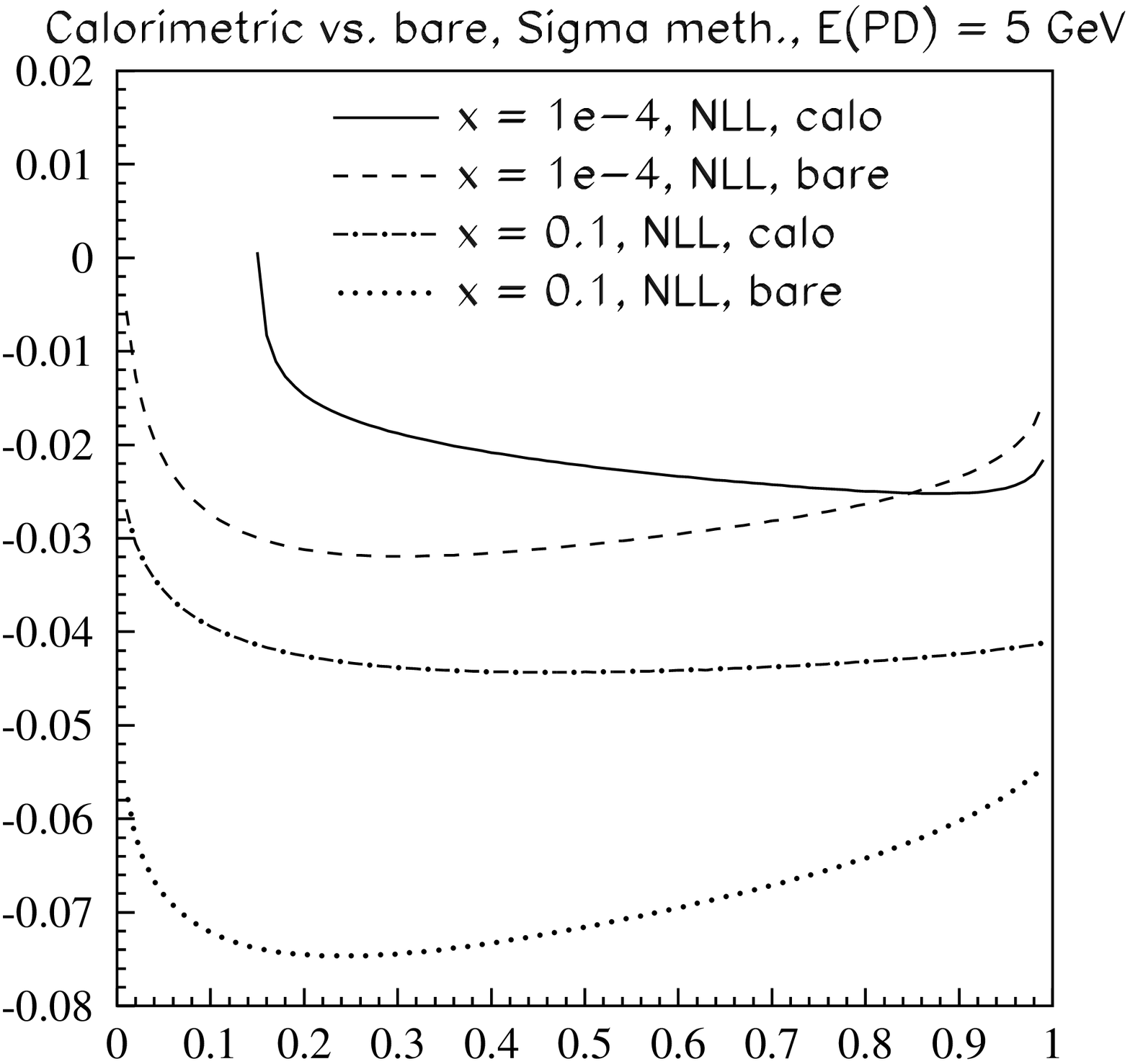}
        }
      \put(104,0){$\hat{y}$}
      \put(-6,103){$\delta_\mathrm{RC}$}
    \end{picture}
    \caption{\textit{Comparison of the corrections for calorimetric
      (``calo'') vs.\ bare measurement of the scattered electron at $\hat{x} =
      10^{-4}$ and $\hat{x} = 0.1$ for a tagged energy of 5\GeV\/ for the
      $\Sigma$ method.}}
    \label{fig:plot3}
  \end{center}
\end{figure}

For this reason we should expect a relatively strong dependence of the
corrections on the experimental selection of the events for the $\Sigma$
method.  This is illustrated in figure~\ref{fig:plot3}, that compares the QED
corrections determined for the calorimetric event selection described above to
a bare measurement of the scattered electron.  Indeed we find a significant
effect especially in the region of larger values of $\hat{x}$, where the
corrections are dominated from soft photon emission.

To conclude, we have reported on the calculations of the QED corrections to
DIS with a tagged photon.  A semi-analytical program that incorporates these
corrections is available on request from one of the authors (H.A.).  Although
no dedicated Monte Carlo event generator has been written yet, the
implementation of the above results should be straightforward.





\end{document}